# On the Performance Evaluation of Encounter-based Worm Interactions Based on Node Characteristics

Sapon Tanachaiwiwat
Department of Electrical Engineering
University of Southern California, CA
tanachai@usc.edu

Ahmed Helmy
Computer and Information Science and Engineering
University of Florida, FL
helmy@ufl.edu

*Abstract:* An encounter-based network is a frequently-disconnected wireless ad-hoc network requiring nearby neighbors to store and forward data utilizing mobility and encounters over time. Using traditional approaches such as gateways or firewalls for deterring worm propagation in encounter-based networks is inappropriate. Because this type of network is highly dynamic and has no specific boundary, a distributed counter-worm mechanism is needed. We propose models for the worm interaction approach that relies upon automated beneficial worm generation to alleviate problems of worm propagation in such networks. We study and analyze the impact of key mobile node characteristics including node cooperation, immunization, on-off behavior on the worm propagations and interactions. We validate our proposed model using extensive simulations. We also find that, in addition to immunization, *cooperation* can *reduce* the level of worm infection. Furthermore, *on-off* behavior linearly impacts only timing aspect but not the overall infection. Using realistic mobile network measurements, we find that encounters are non-uniform, the trends are consistent with the model but the magnitudes are drastically different. Immunization seems to be the most effective in such scenarios. These findings provide insight that we hope would aid to develop counter-worm protocols in future encounter-based networks.

## 1. Introduction

An encounter-based network is a frequently-disconnected wireless ad-hoc networks requiring close proximity of neighbors, i.e., encounter, to disseminate information. Hence, we call this the "encounter-based network" which can be considered as a terrestrial delay-and-disruptive-tolerant network. It is an emerging technology that is suitable for applications in highly dynamic wireless networks.

Most previous work on worm propagation has focused on modeling single worm type in well-connected wired network. However, many new worms are targeting wireless mobile phones. The characteristics of worms in mobile networks are different from random-scan network worms. Worm propagations in mobile networks depend heavily on user encounter patterns. Many of those worms rely on Bluetooth to broadcast their replications to vulnerable phones, e.g., Cabir and ComWar.M [16]. Since Bluetooth radios have very short range around 10-100 meters, the worms need neighbors in close proximity to spread out their replications. Hence, we call this "encounter-based worms". This worm spreading pattern is very similar to spread of packet replications in delay tolerant networks [18, 22], i.e., flooding the copies of messages to all close neighbors. An earlier study in encounter-based networks actually used the term "*epidemic routing*" [18] to describe the similarity of this routing protocol to disease spreading.

Using traditional approaches such as gateways or firewalls for deterring worm propagation in encounter-based networks is inappropriate. Because this type of network is highly dynamic and has no specific boundary, a fully distributed counter-worm mechanism is needed. We propose to investigate the worm interaction approach that relies upon automated beneficial worm generation [1]. This approach uses an automatic generated beneficial worm to terminate malicious worms and patch vulnerable hosts.

Our work is motivated by wars of Internet worms such as the war between NetSky, Bagle and MyDoom [13]. This scenario is described as "worm interactions" in which one or multiple type of worm terminates or patches other types of worms. In [14, 15], we have classified worm interaction types. However, this is the first study on the effect of fundamental characteristics of node behavior on worm propagation.

There are many important node characteristics to be considered, but we focus only a fundamental subset including node cooperation, immunization and *on-off* behavior. We shall show that these are key node characteristics for worm propagation in encounter-based networks. Other characteristics such as trust between users, battery life, energy consumption, and buffer capacity are subject to further study and are out of scope of this paper.

The majority of routing studies in encounter-based networks usually assume ideal node characteristics including full node cooperation and *always-on* behavior. However, in realistic scenarios, nodes do not always cooperate with others and may be *off* most of the time [26]. In worm propagation studies, many works also assume all nodes to be susceptible (i.e., not immune) to worm infection. An *immune* node does not cooperate with

Much of this work was performed at the University of Southern California with support from NSF awards: CAREER 0134650, ACQUIRE 0435505 and Intel.

infected hosts and is not infected. To investigate more realistic scenarios, we propose to study the mobile node characteristics and analyze the impact of node cooperation, immunization and *on-off* behavior on the worm interactions. Cooperation and *on-off* behavior are expected to have impact on the timing of infection. Intuitively, cooperation makes the network more susceptible to worm attacks. Immunization, however, may help reduce overall infection level. This paper examines the validity of these expectations, using the overall infection level and timing of infection as metrics (see Section 3.*C*).

Most worm propagation studies only focus on instantaneous number of infected hosts as a metric. We feel that additional systematic metrics are needed to study worm response mechanisms. We utilize new metrics including total infectives, maximum infectives, total lifespan, average lifespan, time-to-infect-all, and time-to-remove-all to quantify the effectiveness of worm interaction.

In this paper, we try to answer following questions: How can we model this *war of the worms* systemically based on node characteristics including cooperation, immunization, and *on-off* behavior in encounter-based networks? What conditions of node characteristics can alleviate the level of worm infection? This worm interaction model can be extended to support more complicated current and future worm interactions in encounter-based networks. Due to limited space, we only model node characteristics on *aggressive one-sided worm interactions* [14] in which there are two types of worms; beneficial worm and malicious worm. The beneficial worm acts as a predator and can terminate the malicious worm (in this case, the prey!). The predator vaccinates and patches infected hosts and susceptible hosts to prevent infection and re-infections from malicious worm.

Our main contributions in this paper is our proposed new *Worm Interaction Model* focusing on node characteristics in encounter-based networks. We also use new metrics to quantify the effectiveness of worm interactions, and are applicable to study any worm response mechanism. We also provide the first study of worm propagation based on real mobile measurements.

Following is an outline of the rest of the paper. We discuss related work in Section 2. Then, in Section 3, we explain the basic worm interaction model, node-characteristics model, and proposed metrics. Then we analyze worm interactions in both uniform and realistic encounter networks. In Section 4, we conclude our work and discuss the future work.

## 2. Related work

Worm-like message propagation or epidemic routing has been studied for delay tolerant network applications [18, 22]. As in worm propagation, a sender in this routing protocol spreads messages to all nodes in close proximity, and those nodes repeatedly spread the copies of messages until the messages reach a destination, similarly to generic flooding but without producing redundant messages. Performance modeling for epidemic routing in delay tolerant networks [22] based on ODEs is proposed to evaluate the delivery delay, loss probability and power consumption. Also the concept of anti-packet is proposed to stop unnecessary overhead from forwarding extra packets copies after the destination has received the packets. This can be considered as a special case of non-zero delay of aggressive one-sided interaction which we consider in our model.

Epidemic models, a set of ordinary differential equations, were used to describe the contagious disease spread including *SI, SIS, SIR SIRS, SEIR* and *SEIRS* models [3, 9, 17] in which *S, I, E, R* stand for Susceptible, Infected, Exposed and Recovered states, respectively. There's an analogy between computer worm infection and disease spread in that both depend on node's state and encounter pattern. For Internet worms, several worm propagation models have been investigated in earlier work [4, 7, 11]. Few works [8, 12, 14, 15] consider worm interaction among different worm types. Our work, by contrast, focuses on understanding of how we can systemically categorize and model worm propagation based on node characteristics in encounter-based networks.

In [1], the authors suggest modifying existing worms such as Code Red, Slammer and Blaster to terminate the original worm types. The modified code retains portion of the attacking method so it would choose and attack the same set of susceptible hosts. In this paper, we model this as aggressive one-sided worm interaction. Other active defenses, such as automatic patching, are also investigated in [19]. Their work assumes a patch server and overlay network architecture for Internet defense. We provide a mathematical model that can explain the behavior of automatically-generated beneficial worm and automatic patch distribution using one-sided worm interaction in encounter-based networks. Our work aims to understand and evaluate automated worm (with patch) generation but we do not address details of vulnerabilities nor related software engineering techniques to generate patches or worms. Active defense using beneficial worms is also mathematically modeled in [12] which focused on delay-limited worm defense in the Internet.

Effect of Immunization on Internet worms was modeled in [10] based on the *SIR* model.

## 3. Worm Interaction Models and Metrics

*Worm interaction* arises in scenario where one worm terminates other worms. To understand worm interaction, we start by examining the concept of the *predator-prey* relationships in Section *A*. Then, in Section *B*, we introduce the basic concept of worm interaction model and finally we propose new metrics in Section *C*. In Section *D*, we provide basic worm interaction model analysis. Then we introduce concept of node characteristics and node-characteristic-based worm interaction model in Section *E*. Then, in Section *F*, we analyze and compare simulation results between uniform and non-uniform (trace-based) worm interactions.

### A. Predator-Prey Relationships

For every worm interaction type, there are two basic characters: Predator and Prey. The ***Predator***, in our case the beneficial worm, is a worm that terminates and patches against another worm. The ***Prey***, in our case the malicious worm, is a worm that is terminated or patched by another worm[1].

A predator can also be a prey at the same time for some other type of worm. Predator can *vaccinate* a susceptible host, i.e., infect the susceptible host (vaccinated hosts become predator-infected hosts or predator infectives) and apply a patch afterwards to prevent the hosts from prey infection. Manual vaccination, however, is performed by a user or an administrator by applying patches to susceptible hosts.

A *termination* refers to the removal of prey from infected hosts by predator; and such action causes prey infectives to become predator infectives. The removal by a user or an administrator, however, is referred to as *manual removal.* For brevity and clarity, manual vaccination and removal are not considered in this paper.

We choose to use two generic types of interacting worms, *A* and *B*, as our basis throughout the paper. *A* and *B* can assume the role of predator or prey depending on the type of interactions.

### B. Worm Interaction Model

Let *S* be the number of vulnerable hosts that are not yet infected by any worm, i.e. *susceptible* at time *t*. Let $I_A$ and $I_B$ be the number of infected hosts by prey and predator at time *t*, respectively. Assume that each user encounters another random user with constant pair-wise contact rate $\beta$ (probability per unit of time of encounter between any pair) and uniform encounter (every node has equal chance to encounter any other node)[2]. We also assume that node's characteristic including cooperation, immunization and *on-off* behavior does not change after infection from prey or predator. We start with the simple case where every node is cooperative, susceptible and always *on*. The state transition diagram in fig.1 and the susceptible rate and infection rates of prey and predator are given by:

$$\frac{dS}{dt} = -\beta S(I_A + I_B) \tag{1-a}$$

$$\frac{dI_A}{dt} = \beta I_A (S - I_B) \tag{1-b}$$

$$\frac{dI_B}{dt} = \beta (SI_B + I_A I_B)\,. \tag{1-c}$$

[1] Note that in other models the malicious worm may also be a predator, and can be studied similarly. We do not present such study for the lack of space.
[2] This assumption is relaxed later in the paper in the trace-based encounter simulations.

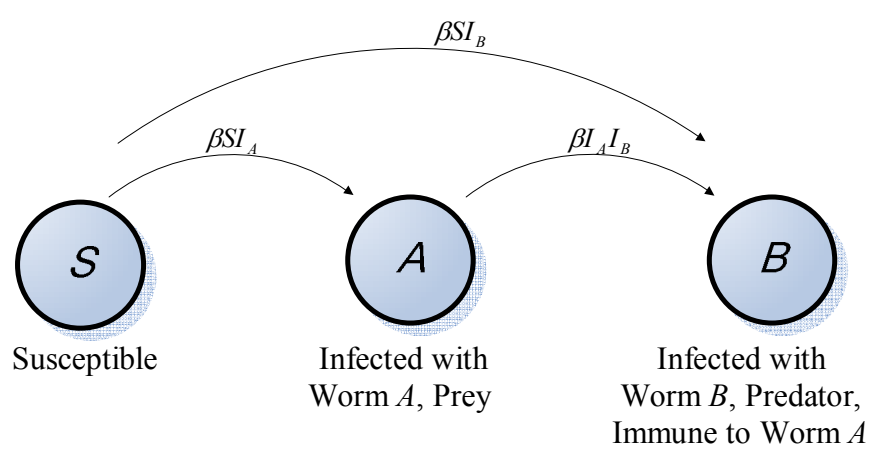

**Figure 1: Aggressive one-sided interaction**

We call this set of equations "*aggressive one-sided interaction model"* where a predator is able to terminate its prey and vaccinate susceptible hosts. We shall vary this model later to capture various node characteristics.

### C. Metrics

To gain insight and better quantify the effectiveness of aggressive one-sided worm interaction, we propose to use the following metrics:

(1) **Total Infectives (*TI*):** the number of hosts ever infected by prey.
(2) **Maximum Infectives (*MI*):** the peak of instantaneous number of prey-infected hosts where $I_A(0) \le MI \le TI$ .
(3) **Total Life Span (*TL*):** the sum of time of individual nodes ever infected by prey. It can be interpreted as the total damage by prey.
(4) **Average Individual Life Span (*AL*):** the average life span of individual prey-infected hosts where $AL \le TL$ .
(5) **Time to Infect All (*TA*):** the time required for predator to infect all susceptible and prey hosts. Its inverse can be interpreted as average predator infection rate.
(6) **Time to Remove All (*TR*):** the time required for predator to terminate all preys where $TR \le TA$ . Its inverse can be interpreted as prey termination rate.

Our goal is to find the conditions to *minimize* these metrics based on node characteristics. We discuss details of node characteristics in Section *E*.

Next we examine the basic worm interaction model and its relationships with above metrics.

### D. Basic Model Analysis

If we want to suppress the initial infection ($\frac{dI_A}{dt}$ =0 at *t*=0), from (1-b), then the required condition for this is

$$I_B(0) = S(0) \tag{2}$$

where $I_B(0)$ and $S(0)$ are the number of prey-infected hosts and susceptible hosts at *t*=0 respectively.

We obtain from this condition that

$$TI = MI = I_A(0),\ I_A(\infty) = 0 \tag{3}$$

where $I_A(\infty)$ is the number of prey-infected hosts at $t=\infty$.

However, we can see from (2) that the threshold can only be obtained by requiring the initial number of predator to be at least equal to number of susceptible hosts (a trivial condition). If that condition cannot be met, i.e., $I_B(0) < S(0)$, then we can only have certain acceptable level of infection and *TI* can be derived from

$$TI = \int_{t=0}^{\infty} \beta S I_A dt \qquad (4)$$

*MI* can be found where $\frac{dI_A}{dt}$ =0 at $t > 0$, in which

$$I_B(t) = S(t) \qquad (5)$$

Let *Y* be the initial infected host ratio which is a ratio of predator initial infected hosts to prey initial infected hosts, i.e., $Y = \frac{I_B(0)}{I_A(0)}$ where $0 < Y < N - S(0)$ and *N* is the total number of nodes in the network.

In figures 2, 3 and 4, we show the metrics characteristics based on *Y* and validate our models through the encounter-level simulations. We simulate and model 1,000 mobile nodes with $\beta =$ 6x10$^{-6}$ sec$^{-1}$, $I_A(0) = 1$, and $1 \le I_B(0), S \le 998$. Each simulation runs at least 1,000 rounds and we plot the mean values for each *Y*.

We assume uniform and constant $\beta$ as well as $I_A(0)$. We adjust *Y* to find the optimal range to minimize our proposed metrics where $Y_{min} = 1$ and $Y_{max} = 998$.

In Fig.2, we show the relationships of *TI* and *MI* as the function of *Y*. *TI* and *MI decrease exponentially as Y increases.* The reason we still keep $I_A(0)$ small is to have wider range of *Y* with the same size of *N*. *MI* (as a fraction of *N*) is more accurately predicted by the model. The ratio of *TI* to *MI* is constant but it gets smaller towards the largest *Y*. We also find that if $S(0):I_B(0):I_A(0)$ is constant then $MI:N$ and $TI:N$ are also constant even *N* changes.

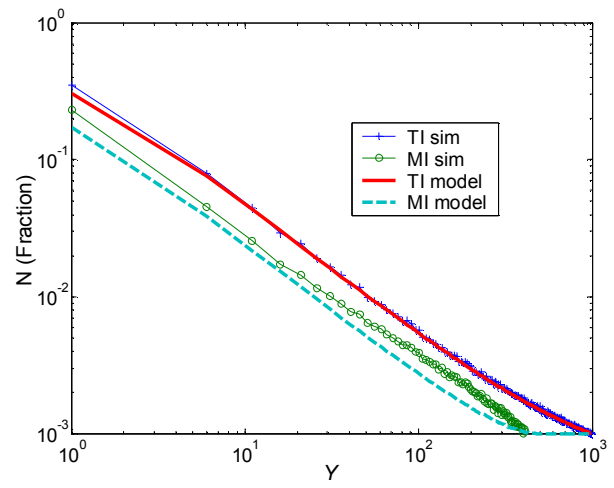

**Figure 2: Relationships of *TI* and *MI* with *Y***

Because *TL* is the accumulated life of individual prey until the last prey has been removed by predator whose duration indicated by *TR*. we can simply derive *TL* based on the numerical solutions from (1-b) as follows:

$$TL = \sum_{t=o}^{\infty} I_{A_t} \Delta t \qquad (6)$$

Since *AL* is the average life span for each node that has been terminated by predator which is equal to the number of nodes that are ever infected, *AL* can be derived from

$$AL = \frac{TL}{\int_{t=0}^{\infty} \beta I_A I_B dt} = \frac{TL}{\int_{t=0}^{\infty} \beta S I_A dt} = \frac{TL}{TI} \qquad (7)$$

*TL* and *AL* trends are mostly accurately predicted by the model. The *AL* errors are due to the errors of estimated *TL*.

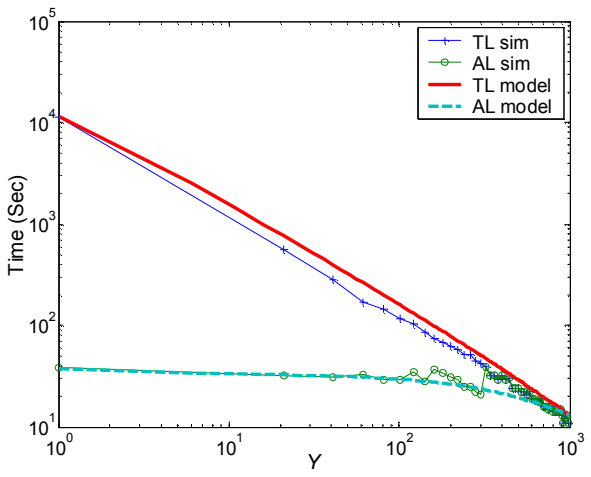

**Figure 3: Relationships of *TL* and *AL* with *Y***

From Fig.3, *TL decreases exponentially as Y increases. AL,* on the other hand*,* is almost constant for all *Y*. It is interesting to see that *TL* and *AL* are merging at their minimum when $Y = Y_{max}$. As we can see that $TL_{min}$ and $AL_{min}$ do not reach zero at $Y_{max}$ because the next encounter time of a prey-infected host with *any* of predator-infected host ($I_B(0)$) requires average of $1/I_B(0)\beta$. Furthermore, from (7), $TL_{min} = TI_{min}AL_{min}$, $TL_{min}$ and $AL_{min}$ merge to each other because $TI_{min} = I_A(0) = 1$.

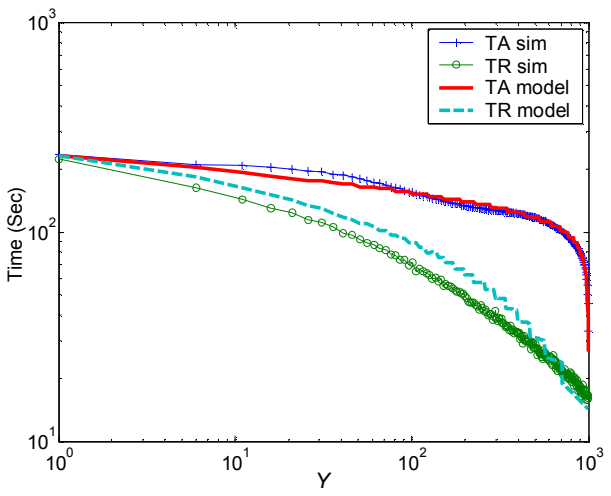

**Figure 4: Relationships of *TA* and *TR* with *Y***

From the observation in Fig.4, *TR* reduces much faster than *TA* with the increase of *Y*. *TR* decreases exponentially as *Y increases*. *TA* starts to be reduced rapidly when $Y \approx Y_{max}$. At $Y_{max,}$ we can see that $TA_{min}=TR_{min}=AL_{min}$,

Note that *TA* is also similar to the average time for every node to receive a copy of a message from a random source in an encounter-based network which can be derived as $(2\ln N + 0.5772)/N\beta$ [24].

### *E. Node characteristics*

Earlier we assume that all nodes are *fully cooperative*, *susceptible* to both prey and predator and *"always-on"* in Section *B*., and hence each encounter guarantees a successful message (worm) transfer.

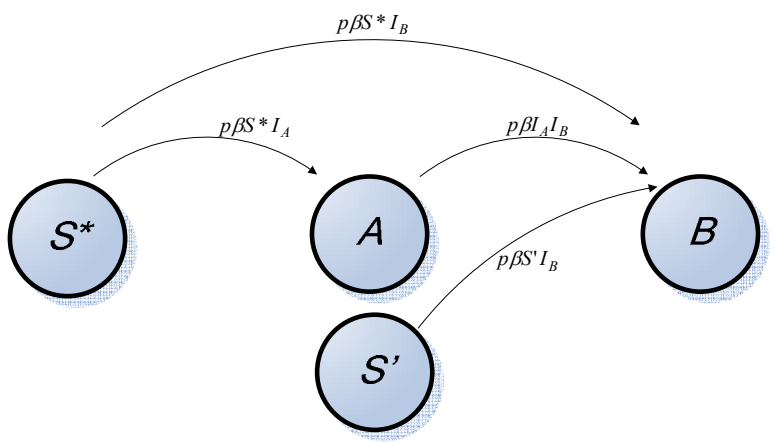

**Figure 5: Aggressive one-sided interaction with node characteristics**

In this section, we investigate the scenarios that do not follow above assumptions regarding these three important node characteristics. We assume these characteristics are consistent through out its life time of the networks.

**(1) Cooperation:**

Cooperation is the willingness of node to forward the message (worm) for other nodes. The opposite characteristic is known as selfishness. Intuitively, cooperation may seem to make the network more vulnerable. However, unlike immunization, cooperation is expected to equally slow down both prey and predator propagations. Hence, the effect of cooperation is hard to anticipate.

**(2) Immunization:**

Not all nodes are susceptible to the prey either because of their heterogeneous operating systems and their differences of promptness to remove the vulnerability from their machines. Hence partial of nodes can be immune to prey and will slow down the overall prey infection. It is expected to improve the overall targeted metrics that we mention earlier. Because even some nodes are immune to the prey but they still help forwarding the predator to other nodes and it is expected to have no positive impact on *AL*, *TA* but reduce *TL* and *TR* simply because of less number of nodes to be removed.

**(3) *On-off* behavior:**

A node is able to accept or forward the packet based on the *on-off* characteristics. In reality, devices are "*on*" or active only a fraction of the time. Activity may be related to mobility. For instance, a mobile phone is usually *on*, while lap top is unlikely to be mobile while *on*[3]. We model the transition from *on* to *off*, and vice versa, probabilistically. The probability is determined at the beginning of each time interval. Hence the contact rate is expected to be proportionally reduced according to the probability that the node cannot forward or accept the packets because of *on-off* status.

Let $c$ be the fraction of $N$ that are willing to be *cooperative* where $0 \le c \le 1$ and $N$ is the total number of nodes in the networks. Let $i$ be the fraction of cooperative nodes that are *immune* to prey where $0 \le i \le 1$. We assume that initial predator and prey hosts are cooperative then the number of susceptible hosts for both prey and predator is $S^*$ where $S^*(0) = c(1-i)N - I_A(0)$ and number of susceptible hosts for predator only is $S'$, where $S'(0) = ciN - I_B(0)$. Note that $N = S^* + S' + I_A + I_B$ and $S = S^* + S'$. We define the probability of "*on*" behavior as $p$ and "*off*" behavior as 1-$p$ where $0 \le p \le 1$. Hence contact rate for both predator and prey is $p\beta$.

Based on these definitions, the node-characteristic-based aggressive one-sided model can be shown as follows:

$$\frac{dS^*}{dt} = -p\beta S^*(I_A + I_B) \quad \text{(8-a)}$$

$$\frac{dS'}{dt} = -p\beta S' I_B \quad \text{(8-b)}$$

$$\frac{dI_A}{dt} = p\beta I_A(S^* - I_B) \quad \text{(8-c)}$$

$$\frac{dI_B}{dt} = p\beta((S^* + S')I_B + I_A I_B) \quad \text{(8-d)}$$

Similarly to Section *D,* We use this model to derive metrics that we are interested. The differences between the conditions of this model and that of basic model to minimize the metrics are investigated here.

If we want to suppress the prey initial infection, then we need

$$I_B(0) = S^*(0) \quad (9)$$

Assume small $I_A(0)$ and $I_B(0)$ when compared with $N$, hence $S^*(0) \approx c(1-i)S(0)$; required $I_B(0)$ to stop prey initial infection is therefore also reduced approximately by the factor of $c(1-i)$ when compared with (2). *TI,* similarly derived to (4), is

$$TI = p\int_{t=0}^{\infty} \beta S^* I_A dt \quad (10)$$

As contact rate is changed due to *on-off* behavior, *TA* which $Y = 1$, can be derived as follows:

$$TA = (2\ln N + 0.5772)/pN\beta \quad (11)$$

Our model can also be used to model node-characteristic-based one-worm-type propagation which equivalent to epidemic routing by assigning $I_B(0) = 0$ or $I_A(0) = 0$ in (8-a) to (8-d).

### *F. Simulation results*

In this section, we start by validating our models with uniform-encounter simulation. Then, we compare the relationships of node characteristic with our proposed metrics in uniform and non-uniform (trace-based) encounter networks.

[3] This is observed from measurements [25, 26] and is captured in our study using trace-driven simulations.

### (1) Uniform Encounters

We use encounter-level simulations to simulate uniform encounter of 1,000 mobile nodes with $\beta$ = 6x10$^{-6}$ sec$^{-1}$, and $I_A(0) = I_B(0) = 1$. Each simulation runs 10,000 rounds and we plot the median values for each *i*. The lag time between predator and prey initial infection is 0 sec. We vary cooperation (*c*) from 20% to 100%, immunization (*i*) from 0% to 90% with 100% "*on*" time for the first part of experiments (Fig. 6(a)-(f)) and we vary "*on*" time from 10% to 90% with 90% cooperation and 10% immunization, for the second part (Fig.6(g)-(h)). The first part aims to analyze the impact of cooperation and immunization on worm interaction whereas the second part aims to analyze the *on-off* behavior.

From fig. 6 (a)-(f) we find that increase of cooperation, surprisingly, reduces malicious worm infection for all the metrics. (Note that increase of cooperation actually increases absolute *TI* and absolute *MI*, but relative *TI* (or $TI/N^*$) and relative *MI* (or $MI/N^*$) are reduced where number of cooperative-susceptible nodes $N^* = c(1-i)N$ ).

Similarly, for immunization fig. 6 (a)-(f) shows that immunization reduces all categories of metrics except *AL*. With the increase of immunization, *TI* is reduced much faster than *TL,* thus increase of immunization increases *AL*. Furthermore, increase of immunization, as expected, reduces *TR* because of less number of possible prey-infected hosts.

*Cooperation reduces AL and TR significantly* than it does to other metrics. *Immunization, however, reduces relative TI, relative MI and TL more significantly* than it does other metrics. With equal increase (20% to 80%), immunization at cooperation = 100% reduces relative *TI*, relative *MI* and *TL* approximately 8.8 times, 2.7 times, and 10.6 times ,respectively, more than cooperation does at immunization = 0%. On the other hand, cooperation reduces *TR* approximately 3.3 times more than immunization does. As shown in fig. 6(e), unlike immunization, only cooperation can reduce *TA*.

The impact of *on-off* behavior (*p*) is clear in fig. 6 (g) and (h). As expected, with variant of *"on" time*, there is no difference in relative *TI* and relative *MI*. The ratio of contact rate between predator and prey is an indicator of the fraction of infected hosts irrespective of the contact rate. In this case, the ratio of contact rate is always 1.0, and hence the constant of relative *TI* and relative *MI*.

*TL*, *AL TA* and *TR exponentially decrease* with the increase of *"on" time* causing reduction of inter-encounter time. Our model shows a good agreement with simulation results for most of the scenarios based on node characteristics.

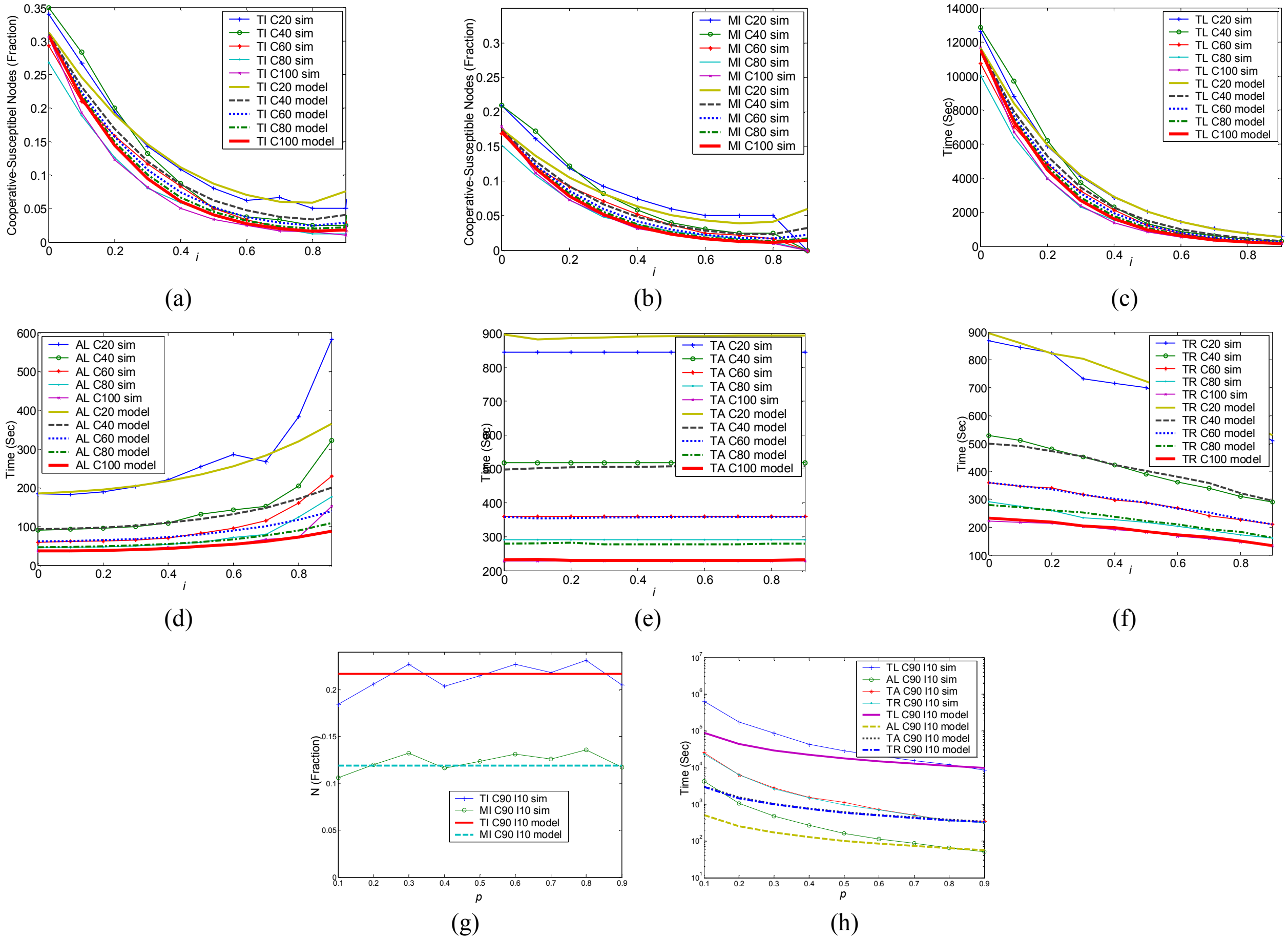

**Figure 6: Effects of cooperation (*c*), immunization (*i*) and *on-off* behavior (*p*) on uniform-encounter worm interactions**

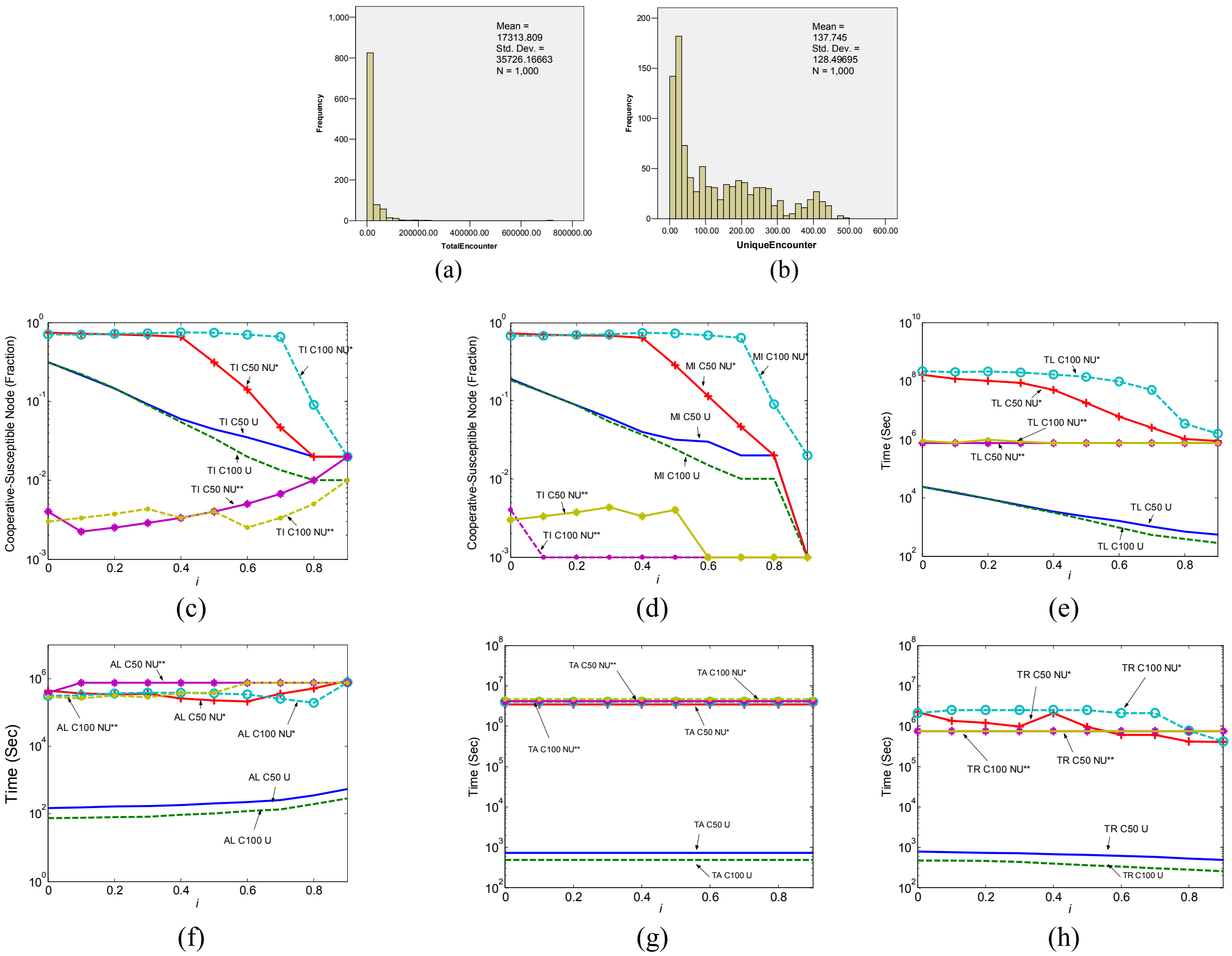

**Fig. 7 Trace-based statistics and simulation results: histograms of (a) total encounter/node and (b) unique encounter/node, and effects on cooperation (*c*), immunization (*i*) and *on-off* behavior (*p*) on (c)*TI* and *MI* (d) *TL* and *AL* and (e) *TR* in non-uniform-encounter worm interaction (U: Uniform, NU: Non-uniform, *: contact rate of initial prey is higher, **: contact rate of initial predator is higher)**

### (2) Non-uniform Encounters

We investigate the consistency of the model-based results with those generated using measurement-based real encounters. We drive our encounter-level simulations using the wireless network traces of the University of Southern California of 62 days in spring 2006 semester[25]. We define an encounter as two nodes sharing the same access point at the same time. We randomly choose 1,000 random nodes from 5,000 most active nodes based on their online time from the trace. Their median $\beta$ is 1.2x10$^{-6}$ sec$^{-1}$and median number of unique encounter node is 94. We use $I_A(0)=1$ and $I_B(t)=1$ where $t$ is the delay between initial predator-infected host and initial prey-infected host in the simulation. This delay was introduced as the traced delay between the two groups in which one group for initial predator-infected host and another for initial prey-infected-host. Each group accounts for 3% of total population. The first group has average contact rate $\beta$=2.7x10$^{-6}$ sec$^{-1}$, and the second group has average contact rate $\beta$=3.6x10$^{-6}$ sec$^{-1}$. When contact rate of the initial predator-infected host is higher than that of the initial prey-infected host, we call this scenario "*Fast predator*". On the other hand, when contact rate of initial predator-infected host is lower than that of prey, we call this scenario "*Slow predator*". From the trace, the median arrival delay between initial predator-infected host and initial prey-infected host is -539,795 sec (6.25 days) for *"Fast predator"*, and 539,795 sec for "*Slow predator*". For comparison between uniform and non-uniform encounter, we directly add the plot of metrics from encounter-level simulation of worm interaction in uniform encounter networks with the same contact rate ($\beta$=1.2x10$^{-6}$ sec$^{-1}$) and the same number of nodes with arrival delay = 0 sec.

In fig. 7, we find that immunization (*i*) is still a very important factor to reduce relative *TI*, relative *MI*, *TL*, and *TR*. However, unlike uniform-encounter worm interaction, we find that higher cooperation does not necessarily help reduce relative *TI*, relative *MI*, *TL*, *AL* and *TR*.

We believe that because of non-uniform encounter patterns (as shown in fig. 7(a)-(b)) and significant lag time between an initial prey-infected host and an initial predator-infected host, there are several differences of the metrics with uniform and non-uniform encounter networks. The main reasons of non-uniform contact rate and non-uniform number of unique contact users are non-uniform *on-off behavior and location preferences*. From [23], there were heavy and light users based on their online time, we know that only 50% of users were online more than 20% of the whole semester. In fig. 7(a)-(b), we find that user's encounter in the trace is highly skewed, i.e., top 20% of user's total encounter account for 72% of all users' encounter and 70% of users encounter less than 20% of total unique users.

Hence, the metrics of worm interaction in non-uniform encounter networks in fig. 7 deviate greatly from the results from that of uniform encounter networks. In fig. 7(c)-(d), relative *TI* and relative *MI* with "*Slow predator*" is much worse than that of uniform encounter networks. On the other hand, the significant improvement of relative *TI* and relative *MI* are shown with "*Fast predator*". In fig. 7(e), *TL* with "*Fast predator*" is almost two orders of magnitude lower than *TL* with "*Slow predator*" but still much higher than *TL* of uniform encounter networks. However, as shown in fig. 7(e)-(f), *AL* with "*Fast predator*" has not shown significant differences than *AL* with "*Slow predator*".

### 4. Summary and Future Work

In this paper, we propose a node-characteristics-based model and metrics as a performance evaluation framework for worm interactions in encounter-based networks, with focus on cooperation, immunization, and "on-off" behavior. We find that in uniform encounter networks, immunization is the most influential characteristics for total infectives, maximum infectives and total life span. Cooperation and on-off behaviors greatly affect average individual life span, time-to-infect-all and time-to-remove-all. Our model also shows a very good agreement with uniform-encounter simulation results.

Based on realistic mobile networks measurements, we find that the contact rate and the number of unique encounters of users are not uniform. This causes worm infection behavior to deviate significantly from that of uniform encounter networks, even though the general trends remain similar to the model.

In addition, the level of infection is now determined by the contact rate of the initial predator and prey-infected hosts. A higher contact rate of initial predator (than prey) infected hosts significantly reduces the total infectives and maximum infectives when compared to those of the opposite scenario.

In such networks, immunization seems to be more important factor than cooperation. Hence, enforcing early immunization and having mechanism to find a high-contact-rate node to use as an initial predator-infected host is critical to contain worm propagation in encounter-based networks. We believe that node-characteristics model for uniform encounter networks can be extended with delay and cluster behavior to explain effect of node characteristics on worm interaction in non-uniform encounter networks of the future.

### References

[1] F. Castaneda, E.C. Sezer, J. Xu, *"WORM vs. WORM: preliminary study of an active counter-attack mechanism"*, ACM workshop on Rapid malcode, 2004

[2] J. C. Frauenthal. *Mathematical Modeling in Epidemiology*. Springer-Verlag, New York,1988

[3] Analysis of the Sapphire Worm - A joint effort of CAIDA, ICSI, Silicon Defense, UC Berkeley EECS and UC San Diego CSE (http://www.caida.org/analysis/security/sapphire

[4] A. Ganesh, L. Massoulie and D. Towsley, *The Effect of Network Topology on the Spread of Epidemics*, in IEEE INFOCOM 2005.

[5] Z. Chen, L. Gao, and K. Kwiat, "*Modeling the Spread of Active Worms*", IEEE INFOCOM 2003

[6] W. O. Kermack and A. G. McKendrick: "*A Contribution to the Mathematical Theory of Epidemics"*. Proceedings of the Royal Society 1997; A115: 700-721.

[7] D. Moore, C. Shannon, G. M. Voelker, and S. Savage, "*Internet Quarantine: Requirements for Containing Self Propagating Code*", in IEEE INFOCOM 2003.

[8] D. M. Nicol, "*Models and Analysis of Active Worm Defense*", Proceeding of Mathematical Methods, Models and Architecture for Computer Networks Security Workshop 2005.

[9] P. Szor, "*The Art of Computer Virus Research and Defense*" (Symantec Press) 2005

[10] S. Tanachaiwiwat, A. Helmy, "*Computer Worm Ecology in Encounter-based Networks (Invited Paper), IEEE Broadnets 2007*

[11] S. Tanachaiwiwat, A. Helmy, "*Modeling and Analysis of Worm Interactions (War of the Worms)*" *IEEE Broadnets* 2007

[12] Trend Micro Annual Virus Report 2004 http://www.trendmicro.com

[13] H. Trottier and P. Phillippe, "*Deterministic Modeling Of Infectious Diseases: Theory And Methods*" The Internet Journal of Infectious Diseases ISSN: 1528-8366

[14] A.Vahdat and D. Becker. *Epidemic routing for partially connected ad hoc networks*. Technical Report CS-2000.

[15] M. Vojnovic and A. J. Ganesh, *"On the Effectiveness of Automatic Patching"* , ACM WORM 2005, The 3rd Workshop on Rapid Malcode, George Mason University, Fairfax, VA, USA, Nov 11, 2005.

[16] X. Zhang, G. Neglia, J. Kurose, and D. Towsley. "*Performance Modeling of Epidemic Routing*", to appear Elsevier Computer Networks journal, 2007

[17] C. C. Zou, W. Gong and D. Towsley, " *Code red worm propagation modeling and analysis*" Proceedings of the 9th ACM CCS 2002

[18] D.E. Cooper, P. Ezhilchelvan, and I. Mitrani, *A Family of Encounter-Based Broadcast Protocols for Mobile Ad-hoc Networks*, In Proceedings of the Wireless Systems and Mobility in Next Generation Internet. 1st International Workshop of the EURO-NGI Network of Excellence, Dagstuhl Castle, Germany, June 7-9 2004 Kotsis, G. and Spaniol, O. (eds.) Lecture Notes in Computer Science Volume 3427 pp. 235-248 Springer 2005

[19] W. Hsu, A. Helmy, "*On Nodal Encounter Patterns in Wireless LAN Traces*", The 2nd IEEE Int.l Workshop on Wireless Network Measurement (WiNMee), April 2006.

[20] W. Hsu, A. Helmy, "*On Modeling User Associations in Wireless LAN Traces on University Campuses*", The 2nd IEEE Int.l Workshop on Wireless Network Measurement (WiNMee), April 2006.